**Vadim R. Madgazin      vadim@vmgames.com**

# Dichotic harmony for the musical practice.

**Abstract.**

The dichotic method of hearing sound adapts in the region of musical harmony. The algorithm of the separation of the being dissonant voices into several separate groups is proposed. For an increase in the pleasantness of chords the different groups of voices are heard out through the different channels of headphones. Is created two demonstration program for PC.

**Keywords:** music, harmony, chord, dichotic listening, dissonance, consonance, headphones, pleasantness, midi.

**Вадим Р. Мадгазин      vadim@vmgames.com**

# Дихотическая гармония для музыкальной практики.

**Абстракт.**

В работе рассмотрена идея применения дихотического способа прослушивания звука в области музыкальной гармонии. Для увеличения благозвучия аккордов предложен алгоритм разделения диссонирующих пар голосов на несколько отдельных не связанных или частично связанных групп с прослушиванием их через головные стереофонические телефоны. На этой основе созданы две демонстрационные программы для ПК.

## Введение.

Как известно для восприятия звуковых сигналов у человека есть два органа слуха, правое и левое ухо. В естественных условиях звуковые сигналы, попадающие в эти два органа как правило сильно коррелированны, однако имеющиеся в них различия позволяют слуховой системе субъекта выделять значительную добавочную информацию, например пространственное положение источников звука.

Если на разные уши подавать разные звуки (не коррелированные друг с другом), то такое слушание называется дихотическим. Противоположным будет случай монофонического (диотического) слушания, когда в оба уха попадает в точности один и тот же сигнал.

Хорошо известно, обработка звуковой информации в нервной системе человека идёт в два этапа. Сначала производится раздельная первичная обработка исходных сигналов каждого уха, затем совместная вторичная обработка сигналов от обоих ушей.
Различные феномены звукового восприятия в разной степени связаны с каждым из этих двух этапов.

Феномен музыкального диссонанса (например "неприятный" звук некоторых сочетаний двух или более музыкальных нот) обязан своим возникновением в основном первому из них.
А феномен эмоционального восприятия звука обязан своим возникновением второму.

Проверить эти факты можно например путём сравнения ощущений при диотическом и дихотическом прослушивании звука музыкальных аккордов, составленных из двух и трёх голосов.

## Диссонанс и консонанс.

Из обширной литературы известно, что диссонантность двух звуков при дихотическом восприятии значительно уменьшается по сравнению с диотическим, см. напр. [1].

Это происходит потому, что при дихотическом слушании два звуковых сигнала подаются на разные уши и они могут взаимодействовать в психике субъекта только опосредованно, на подобии взаимодействия двух нот, взятых последовательно одна за другой. В этом смысле дихотический гармонический интервал имеет некие общие черты с диотическим мелодическим интервалом.

Хотя это может быть и не совсем так, но общепринято мнение, что при уменьшении диссонантности звука его консонантность увеличивается на такую же величину. Следовательно можно утверждать, что при разведении двух диссонирующих звуков по разным ушам они становятся более консонирующими.

Далее мы будем использовать понятия "консонантности" и "благозвучия" аккордов как почти эквивалентные.

В музыкальной гармонии (в обычном, диотическом варианте) существует довольно ограниченное число консонирующих аккордов. Например в пределах октавы есть всего 6 трёхголосных "вполне благозвучных" аккордов. При увеличении числа голосов диссонантность аккордов только увеличивается, так как быстро растёт число всевозможных пар голосов и соответственно вероятность того, что среди них появится хотя бы один диссонирующий интервал.

Однако в диссонантность аккорда могут вносить вклад не только пары голосов. По крайней мере некоторые тройки голосов тоже на это способны, однако их вклад в таковую диссонантность как правило значительно меньше. Например общеизвестно, что состоящие из одних и тех же интервалов минорное и мажорное трезвучия различаются по диссонансности - у минорного она немного больше.

Я считаю, что причиной этому служит так называемая виртуальная высота сложного звука [2]. У любого минорного аккорда виртуальная высота звука меньше и слабее выражена, чем у соответствующего мажорного. Примерно то же самое имеет место у виртуальной высоты диссонирующих пар голосов по сравнению с консонирующими.

Однако при дальнейшем изложении из-за малости дополнительного вклада в общий диссонанс сочетаний трёх и более звуков мы будем полагать что он отсутствует, и это упрощение не повлияет на принципиальные результаты данной работы.

## Дихотическая гармония.

Как представляется "благозвучная" музыка должна по преимуществу использовать "благозвучную" гармонию, т.е. консонирующие аккорды.

Т.о. вполне логично предложить для улучшения благозвучия музыкального произведения использовать дихотический способ прослушивания музыки.

Ведь каждый диссонантный многоголосный аккорд можно сделать консонантным (или хотя бы - менее диссонантным) при помощи разделения его на дихотическую пару двух "менее голосных" аккордов. Например, 6-голосный аккорд можно представить как два 3-х голосных, подаваемых на разные уши.

Давайте проанализируем это предложение подробнее.

Кое-какие проблемы тут могут возникнуть, ведь у разных ушей каждого человека обычно немного разная частотная характеристика чувствительности, а кроме того одно из двух ушей при дихотическом слушании как правило является "доминантным"...

Однако в чём же основное предназначение музыки? Наверное - передавать эмоции. Но когда мы прослушиваем даже монофоническую музыку обычным способом, из-за акустики зала и взаимного расположения источников звука и нашей головы разные уши всё равно слышат сигналы с разными спектрами. И, кроме того, одна и та же музыка может вызывать неодинаковые эмоции у разных людей и даже у одного человека в разное время. И всё это не мешает нам любить музыку, не так ли?

То есть, если даже при дихотическом прослушивании характер воздействия музыки будет отличаться от обычного - это не причина отказываться от тех преимуществ, которые оно даёт: (как правило) значительное улучшение благозвучия музыкальных аккордов при сохранении (в основном) их эмоционального содержания.

Однако по ряду причин нам не стоит ограничиваться дихотическим прослушиванием уже написанной ранее музыки.
Во-первых диссонантность гармонии - это часто не недостаток, а плод осознанного творчества композитора.
Во-вторых дихотическая гармония вовсе не отменяет диссонансы, но она способна значительно расширить количество "компактных" благозвучных аккордов и благодаря ей возможно осмысленное использование намного более много-голосных аккордов.

Таким образом, для современного композитора открываются новые горизонты, пространство благозвучной гармонии примерно удваивается и вообще - усложняется.
Больше пространства - больше музыки, не так ли?

Ниже я буду кратко обозначать пространство дихотической гармонии как 2H (по аналогии с 2D, двумерным геометрическим пространством).

## Музыкальные шкалы.

Естественно, применение 2Н не ограничивается равномерно темперированным строем с 12-ю нотами в октаве (РТС12), её возможности гораздо шире.

Фактически в любом музыкальном строе, с произвольным числом нот в октаве, с равномерными и неравномерными интервалами между нотами можно таким же образом как в РТС12 использовать дихотическое деление аккордов на пары и при этом получать более приятный и чистый звук нежели чем в обычном случае.

Даже при плавном и произвольном изменении высоты одного или двух звуков дихотическое слушание позволяет полностью избавляться от биений. Можно взять два звука переменной частоты и подать их в разные уши. Или можно подать в одно ухо консонантный аккорд, а во второе - звук переменной частоты и это будет довольно красиво.

## Особенности и ограничения 2Н.

К сожалению, или к счастью у дихотической гармонии есть один большой недостаток (см ниже) и ещё один поменьше:
одновременно извлекаемые звуки всех инструментов мелодии должны рассматриваться как один аккорд, который необходимо развести по разным ушам. При этом в ряде случаев звуки некоторых инструментов вынуждены будут "скакать" из одного уха в другое...

Поэтому скорее всего область применения 2Н будет ограничена музыкой для одного или двух разных инструментов.

Но мне кажется, что в классической музыке это скорее преимущество, чем недостаток. Моя самая любимая музыка - это музыка для одного отдельного инструмента - для фортепиано, клавесина, скрипки, органа, вокала без сопровождения.

А вот музыка для двух (и более!) одинаковых инструментов наоборот может только выиграть от использования 2Н.

При звукорежиссуре 2Н музыкальных произведений (например для придания звуку объёма, реверберации) необходимо обрабатывать каждый из двух каналов звука (или трёх точек панорамы) независимо, чтобы никакая часть сигнала одного канала не проникала в другой.

Большой же недостаток состоит в том, что для прослушивания 2Н музыки вам необходимо иметь головные телефоны (или другую аппаратуру для подачи разных звуков на разные уши), и все музыкальные инструменты (кроме одноголосных) должны быть или электронными, или безрезонаторными акустическими, адаптированными для снятия звука в виде электрических сигналов (наподобие электро- гитары, скрипки, рояля).

Но в наше время даже чисто электронные музыкальные инструменты достигли такого уровня, что могут отлично имитировать звук практически любого акустического инструмента.

Хотя 2H музыка может быть прослушана в "оригинальном" качестве дома или в любом другом месте, но всё же этот факт не обязательно отменяет возможность и целесообразность посещать 2H-концерт в большом зале с живыми исполнителями.

Дихотическая гармония может претендовать на титул первого и пока единственного феномена "электронной эры", который потенциально способен войти в учебники по классической музыке. Но сначала её должно проверить и одобрить весьма консервативное сообщество профессиональных композиторов и музыкантов.

Но довольно говорить о 2H-теории, давайте начнём переходить к 2H-практике.

## Алгоритмы и программы.

Мною созданы две программы, реализующие и демонстрирующие дихотическую гармонию.

Прежде чем вы приступите к оценке 2H вам необходимо иметь IBM PC совместимый компьютер (ПК) и достаточно качественные стерео наушники, опционально вам может понадобиться подключённый к ПК midi синтезатор.

Звук ПК не должен воспроизводиться через какую-либо акустическую систему, но только через наушники!

Демонстрационные программы Dichotic Harmony Keyboard (DHK ver 1.2) и Dichotic Harmony Accords Generator (DHAG ver 2.03) могут быть скачаны со страницы [3], после чего следует запустить их на установку.

Сначала опишу работу программу DHK. По-видимому на некоторых компьютерах она может быть не вполне работоспособна без подключённого внешнего midi синтезатора.
Сначала прочтите её документацию, весьма краткая версия которой приведена ниже.

"Идея программы состоит в том, чтобы при помощи индивидуального панорамирования голосов аккорда уменьшать его эффективную диссонантность.

Входной midi поток данных складывается с midi потоком от внутреннего генератора аккордов, и направляется в блок анализатора диссонансов аккорда, где в зависимости от ситуации производится то или иное панорамирование каждого отдельного голоса аккорда, т.е. распределение его по правому и/или левому стерео каналу звука."

В DHK используется упрощенная модель расчёта суммарного диссонанса аккорда:
-считается, что диссонанс возникает только между каждой парой голосов, полностью или частично попадающих в один и тот же канал вывода звука;
-диссонанс дихотической пары голосов (попадающих в разные каналы) равен нулю;
-при одинаковой громкости пары голосов их диссонанс определяется по элементам массива диссонансов каждого интервала между голосами в полутонах шкалы PTC12, где интервал служит индексом этого элемента Dissonance[interval];
-при разной громкости голосов диссонанс умножается на отношение громкости слабого голоса к сильному, в частности если один голос в канале в 2 раза слабее другого, то и диссонанс их интервала в 2 раза меньше;
-общий диссонанс аккорда определяется как сумма диссонансов всех пар голосов, целиком сосредоточенных в левом канале звука плюс то же для правого канала звука, а для 3-х точечного аккорда плюс то же для голосов в центре панорамы плюс сумма кросс-дисонансов от взаимодействия каждого голоса из центра панорамы с каждым голосом на краях панорамы.

Наиболее тонкий (и отчасти спорный) момент - выбор числовых значений массива диссонансов. В программе этот массив составлен эмпирически на основе данных из [4] в виде чётных целых чисел, причём для интервалов, превышающих октаву используется простое правило: каждая лишняя октава увеличивает диссонанс интервала на 2. Т.к. диссонанс унисона в DHK не используется, то диссонанс октавы принят за 0.

Вот массив диссонансов интервалов от малой секунды до октавы (1...12 полутонов):
22, 16, 10, 6, 4, 18, 2, 8, 12, 14, 20, 0.

Первоначально я предполагал разводить по разным каналам (дихотически) только те аккорды, суммарный диссонанс которых превышает некоторый заранее заданный порог, а остальные (консонантные) аккорды подавать одинаково на оба канала (диотически). Причём, "оптимальное" значение этого порога в общем случае могло зависеть от количества голосов аккорда. Для реализации этой идеи предназначался набор регуляторов порогов "Dissonance threshold for 2...6 voices".

Однако позже практика показала что наилучшим решением будет выводить дихотически все аккорды, а не только диссонантные. Т.о. в дальнейшем при рассмотрении работы программы DHK все эти пороги должны быть равны нулю.

Ещё один важный момент - количество точек панорамы, по которому в DHK производится расстановка голосов аккорда.
Не для каждого аккорда наилучшим практическим решением является разведение каждого голоса только по одному из двух каналов звука. Для многих 6-голосных аккордов неплохим будет деление их на две 3-х голосные пары. А вот для 5, 4 и особенно для 3-х голосных аккордов это не так!

2Н вариант с 2-мя голосами в одном канале и 1-м в другом - слишком уменьшает "объём" аккорда (всего 1 пара взаимодействующих голосов вместо 3-х пар) и делает каналы неравноправными (1 пара против 0 пар).

Поэтому для 3-х голосных аккордов (и некоторых других) наилучшим решением будет добавление 3-й виртуальной точки в центре панорамы, для которой громкость голоса, находящегося в этой точке делиться одинаковым образом по обоим каналам звука.

Т.о. в DHK используется 1 точка панорамы (в центре) для диотического слушания и 2-3 точки для дихотического: 2 точки по краям (левому и правому), и 3-я точка - в центре панорамы.

Оптимальное звучание 3-х голосных аккордов получается при направлении 3-х разных голосов в 3 разные точки панорамы. Т.к. голос в центре панорамы делится пополам на оба канала звука, то он порождает кросс-диссонанс с каждым отдельным голосом, расположенным целиком по краям панорамы, однако как указано выше этот диссонанс вдвое меньше обычного.

## Сокращенное обозначение аккордов.

В DHK используется специальный способ обозначения аккордов в шкале PTC12, который однозначно связан с составом нот аккорда и охватывает всё множество аккордов. Для этого используется идентификатор аккорда вида NvA, где A порядковый номер аккорда с числом голосов N.

Аккорд номер 1 состоит из N голосов, каждый из которых сдвинут на 1 полутон от другого, это самый плотный аккорд по расположению голосов. Номер 2 отличается от 1 тем, что верхний голос сдвинут еще на 1 полутон вверх.

В общем случае пронумеруем голоса аккорда в порядке увеличения их высоты от 1 до N. При увеличении номера аккорда на 1 происходит сдвиг на 1 полутон вверх (N-1)-го голоса, но так, чтобы он не совпал по высоте с N-м голосом.
Если это невозможно, то производится сдвиг (N-2)-го голоса, а голос (N-1) опускается по высоте до нового уровня (N-2)-го голоса плюс 1 полутон.
Если это невозможно, то производится сдвиг (N-3)-го голоса, а все верхние кроме N-го опускаются к нему и т.д. пока возможности для сдвига "внутренних" голосов аккорда (со 2-го по (N-1)-й) не будут исчерпаны, и тогда производится сдвиг N-го голоса на 1 полутон вверх.

В секции генератора аккорда есть отдельные регуляторы N и A, а также индикатор обозначения аккорда и конкретного состава голосов в виде номеров нот в PTC12:
нота 0 это (например) До 1-й октавы, нота 1 это До-диез, нота 2 это Ре и т.д.

## Начинаем слушать.

Легче всего понять преимущества дихотической гармонии над обычной если сравнить различные аккорды при 2-3-точечном и 1-точечном панорамировании соответственно.

Вот яркий пример. Сыграйте одноголосную гамму с присоединённым дополнительным голосом, сдвинутым на 1 полутон выше основного. Или прослушайте генератор аккорда 2v1 изменяя midi номер ноты 0. В 1-точечном варианте это почти невозможно слушать, так велик диссонанс малой секунды. А вот в 2-точечном варианте эта гамма звучит чуть ли не великолепно!

А теперь прослушайте аккорды 3v8 и 3v9 - в 3-х точечном варианте они заметно "лучше", благозвучнее чем в 1-точечном.

Минорный аккорд 3v18 также выигрывает от перевода в 3-х точечную версию, а вот для мажорного 3v19 почти ничего не меняется, за исключением некоторого уменьшения его "объёма".

Да, при переводе аккордов из 1- в 3-х точечный режим происходит уменьшение "объёма", наполненности, степени соединения голосов в аккорде. Но по моему мнению это с лихвой компенсируется во первых увеличением их благозвучия (кроме мажорных аккордов, для которых оно не меняется), а во вторых тем, что все аккорды становятся как бы более "чёткими", одинаково ясными для восприятия.

Я написал некую субъективную последовательность 3-х голосных аккордов по степени уменьшения их благозвучия. Используя кнопку "Number in accords chain" и меняя номер можно убедиться, что граница сопоставимой благозвучности аккордов в 3-х точечном режиме примерно вдвое дальше от начала последовательности, чем в 1-точечном (на мой слух получается около 37 аккордов 2H против примерно 20 обычных).

На основе личных наблюдений мной составлена таблица характеристик всех 3-х голосных аккордов, расположенных в пределах одной октавы (см. Приложение). Приведу ниже свои комментарии к ней.

Хотя все определяемые на слух цифры таблицы исключительно субъективны, но качественную зависимость они передают скорее всего правильно.

Мой слух различает относительные сдвиги высоты звука примерно в 10 центов, однако пороги чувствительности разных ушей по уровню громкости звука неодинаковы.

Из таблицы хорошо видно, что благозвучие всех аккордов увеличивается в 2H варианте. Т.к. 2-х точечный режим отличается слишком низким уровнем взаимодействия голосов ("Синергией"), он реально не может быть рекомендован к использованию, и благозвучие в этом режиме я вообще не измерял.

Параметр DDiss определялся относительно 1-точечного режима, а вот его "единица измерения" была настолько непостоянна, что сравнение разных аккордов друг с другом по DDiss не имеет смысла. К другим измеренным параметрам данное замечание относится в значительно меньшей степени.

Однако из-за фактора утомления ("притупления слуха") конкретные цифры в далеко отстоящих строках таблицы также могут быть менее сопоставимы по абсолютным величинам, чем стоящие в соседних строках.

Интересный 2H феномен - хорошо заметное различие звука одного и того же аккорда при обмене местами левого и правого каналов звука. Изменения звука похожи на тембровые. Как видно степень выраженности этого эффекта различна для разных аккордов. Причём, как мне представляется, нельзя объяснить данный феномен только фактом разной чувствительности ушей субъекта. Возможно какие-то особенности восприятия дихотического сигнала также имеют место, например - факт доминантности одного из ушей.

## Сложные аккорды.

Программа DHAG позволяет составлять 1...15 голосные аккорды во всём диапазоне PTC12, присваивая каждому голосу одну из трёх точек стерео панорамы: левый край, центр, правый край. Выходом программы является поток midi сообщений, посылаемых на устройство ОС Windows "MIDI Output Device", которое и синтезирует звуковой сигнал.

С помощью этой второй программы возможно конструировать и прослушивать гораздо более сложные дихотические аккорды, а также их довольно длинные последовательности (цепочки до 5000 аккордов, фактически музыку).

Прочтите полностью документацию к DHAG (текст которой можно рассматривать как продолжение данной статьи) и прослушайте все демонстрационные примеры (цепочек) аккордов.

Не забудьте, что все эти демонстрации лишь в малой степени могут иллюстрировать возможности применения дихотической гармонии для музыкальной практики. Настоящая работа, и прежде всего для новаторов-композиторов - впереди!

Все заинтересованные в развитии дихотической гармонии приглашаются к конструктивной дискуссии, email автора указан в начале статьи.

## Ссылки.

**Приложение.**

Таблица характеристик всех 3-х голосных аккордов, расположенных в пределах одной октавы.

Настройки регуляторов DHK.

General Midi Instrument: 020 Reed Organ
Max num voices: 3
Voices panorame sort mode: Increase Sort
Dissonance threshold for 3 voices: 0
Swap left/right speakers: unchecked (меняем это только для измерения Различия)
Dissonance[interval]: 0[0],22[1],16[2],10[3],6[4],4[5],18[6],2[7],8[8],12[9],14[10],20[11]
Number of voices: 3
Midi number of zero note: 60

Расшифровка столбцов таблицы.

"Аккорд" - идентификатор аккорда в формате NvA
"Состав" - номера нот аккорда, ноты без знака звучат в центре панорамы,
со знаком "-" слева,
со знаком "+" справа
"PPN" - Panorame position's number, количество точек панорамы
"TDiss" - Dichotic total dissonance, дихотический суммарный диссонанс
"Благозвучие" - приятность на слух, красота аккорда
"DDiss" - изменение диссонанса при сравнении режима 2,3 точек панорамы с 1-точечным
"Синергия" - объём, наполненность, степень соединения голосов в аккорде
"Различие" - степень отличия при перестановке левого и правого каналов звука.

| Аккорд | Состав | PPN | TDiss | Благозв. | DDiss | Синергия | Различие | Тип аккорда |
|---|---|---|---|---|---|---|---|---|
| 3v1 | 0,1,2 | 1 | 60 | 0 | 0 | 0 | 0 | |
| | 0-,2,1+ | 3 | 19 | 1 | -1 | -1 | 1 | |
| | 1-,0+,2+ | 2 | 16 | | -3 | -3 | | |
| 3v2 | 0,1,3 | 1 | 48 | 0 | 0 | 0 | 0 | |
| | 0-,3,1+ | 3 | 13 | 1 | -2 | -1 | 1 | |
| | 1-,0+,3+ | 2 | 10 | | -4 | -3 | | |
| 3v3 | 0,2,3 | 1 | 48 | 0 | 0 | 0 | 0 | |
| | 2-,0,3+ | 3 | 13 | 1 | -3 | -1 | 2 | |
| | 0-,3-,2+ | 2 | 10 | | -4 | -3 | | |
| 3v4 | 0,1,4 | 1 | 38 | 1 | 0 | 0 | 0 | |
| | 0-,4,1+ | 3 | 8 | 2 | -3 | -1 | 1 | |
| | 1-,0+,4+ | 2 | 6 | | -5 | -3 | | |
| 3v5 | 0,2,4 | 1 | 38 | 2 | 0 | 0 | 0 | секундовый |
| | 0-,4,2+ | 3 | 11 | 3 | -2 | -1 | 2 | |
| | 2-,0+,4+ | 2 | 6 | | -5 | -3 | | |
| 3v6 | 0,3,4 | 1 | 38 | 1 | 0 | 0 | 0 | |
| | 3-,0,4+ | 3 | 8 | 2 | -2 | -1 | 1 | |
| | 0-,4-,3+ | 2 | 6 | | -4 | -3 | | |
| 3v7 | 0,1,5 | 1 | 32 | 1 | 0 | 0 | 0 | |
| | 0-,5,1+ | 3 | 5 | 2 | -2 | -1 | 2 | |
| | 1-,0+,5+ | 2 | 4 | | -5 | -3 | | |
| 3v8 | 0,2,5 | 1 | 30 | 2 | 0 | 0 | 0 | |
| | 0-,5,2+ | 3 | 7 | 4 | -2 | -1 | 2 | |
| | 2-,0+,5+ | 2 | 4 | | -5 | -3 | | |
| 3v9 | 0,3,5 | 1 | 30 | 2 | 0 | 0 | 0 | |
| | 3-,0,5+ | 3 | 7 | 4 | -2 | -1 | 2 | |
| | 0-,5-,3+ | 2 | 4 | | -4 | -3 | | |
| 3v10 | 0,4,5 | 1 | 32 | 1 | 0 | 0 | 0 | |
| | 4-,0,5+ | 3 | 5 | 2 | -3 | -1 | 2 | |
| | 0-,5-,4+ | 2 | 4 | | -4 | -2 | | |
| 3v11 | 0,1,6 | 1 | 44 | 0 | 0 | 0 | 0 | |
| | 0-,6,1+ | 3 | 11 | 1 | -2 | -1 | 1 | |
| | 0-,1+,6+ | 2 | 4 | | -4 | -3 | | |
| 3v12 | 0,2,6 | 1 | 40 | 2 | 0 | 0 | 0 | |
| | 0-,2,6+ | 3 | 11 | 3 | -3 | -1 | 3 | |
| | 0-,2+,6+ | 2 | 6 | | -4 | -2 | | |
| 3v13 | 0,3,6 | 1 | 38 | 1 | 0 | 0 | 0 | уменьшенный |
| | 0-,3,6+ | 3 | 10 | 2 | -3 | -1 | 3 | |
| | 0-,3-,6+ | 2 | 10 | | -3 | -2 | | |
| 3v14 | 0,4,6 | 1 | 40 | 1 | 0 | 0 | 0 | |
| | 0-,4,6+ | 3 | 11 | 2 | -2 | -1 | 3 | |
| | 0-,4-,6+ | 2 | 6 | | -3 | -2 | | |
| 3v15 | 0,5,6 | 1 | 44 | 0 | 0 | 0 | 0 | |
| | 5-,0,6+ | 3 | 11 | 1 | -2 | -1 | 1 | |
| | 0-,5-,6+ | 2 | 4 | | -3 | -2 | | |
| 3v16 | 0,1,7 | 1 | 42 | 0 | 0 | 0 | 0 | |
| | 0-,7,1+ | 3 | 10 | 1 | -2 | -1 | 1 | |
| | 1-,0+,7+ | 2 | 2 | | -3 | -2 | | |
| 3v17 | 0,2,7 | 1 | 22 | 3 | 0 | 0 | 0 | suspended |
| | 0-,7,2+ | 3 | 3 | 4 | -2 | -1 | 2 | задержанный |
| | 2-,0+,7+ | 2 | 2 | | -4 | -3 | | |
| 3v18 | 0,3,7 | 1 | 18 | 4 | 0 | 0 | 0 | минор |
| | 0-,7,3+ | 3 | 4 | 5 | -2 | -1 | 3 | |
| | 3-,0+,7+ | 2 | 2 | | -4 | -3 | | |
| 3v19 | 0,4,7 | 1 | 18 | 5 | 0 | 0 | 0 | мажор |
| | 4-,0,7+ | 3 | 4 | 5 | -1 | -1 | 3 | |
| | 0-,7-,4+ | 2 | 2 | | -2 | -2 | | |
| 3v20 | 0,5,7 | 1 | 22 | 3 | 0 | 0 | 0 | suspended |
| | 5-,0,7+ | 3 | 3 | 5 | -2 | -1 | 2 | задержанный |
| | 0-,7-,5+ | 2 | 2 | | -3 | -2 | | |
| 3v21 | 0,6,7 | 1 | 42 | 1 | 0 | 0 | 0 | |
| | 6-,0,7+ | 3 | 10 | 2 | -2 | -1 | 1 | |
| | 0-,7-,6+ | 2 | 2 | | -3 | -2 | | |
| 3v22 | 0,1,8 | 1 | 32 | 1 | 0 | 0 | 0 | |
| | 0-,8,1+ | 3 | 5 | 2 | -2 | -1 | 1 | |
| | 0-,1+,8+ | 2 | 2 | | -4 | -3 | | |
| 3v23 | 0,2,8 | 1 | 42 | 2 | 0 | 0 | 0 | |
| | 2-,0,8+ | 3 | 12 | 3 | -2 | -1 | 2 | |
| | 2-,0+,8+ | 2 | 8 | | -3 | -2 | | |
| 3v24 | 0,3,8 | 1 | 22 | 5 | 0 | 0 | 0 | мажор |
| | 0-,8,3+ | 3 | 6 | 5 | -1 | -1 | 3 | |
| | 0-,3+,8+ | 2 | 4 | | -3 | -3 | | |
| 3v25 | 0,4,8 | 1 | 20 | 2 | 0 | 0 | 0 | увеличенный |
| | 0-,4,8+ | 3 | 6 | 3 | -2 | -1 | 3 | |
| | 0-,4-,8+ | 2 | 6 | | -3 | -2 | | |
| 3v26 | 0,5,8 | 1 | 22 | 4 | 0 | 0 | 0 | минор |
| | 5-,0,8+ | 3 | 6 | 5 | -2 | -1 | 3 | |
| | 0-,5-,8+ | 2 | 4 | | -3 | -2 | | |
| 3v27 | 0,6,8 | 1 | 42 | 2 | 0 | 0 | 0 | |
| | 0-,8,6+ | 3 | 12 | 3 | -2 | -1 | 2 | |
| | 0-,8-,6+ | 2 | 8 | | -3 | -2 | | |
| 3v28 | 0,7,8 | 1 | 32 | 1 | 0 | 0 | 0 | |
| | 7-,0,8+ | 3 | 5 | 2 | -2 | -1 | 1 | |
| | 0-,7-,8+ | 2 | 2 | | -3 | -2 | | |
| 3v29 | 0,1,9 | 1 | 42 | 1 | 0 | 0 | 0 | |
| | 0-,9,1+ | 3 | 10 | 2 | -2 | -1 | 1 | |
| | 0-,1+,9+ | 2 | 8 | | -4 | -2 | | |
| 3v30 | 0,2,9 | 1 | 30 | 2 | 0 | 0 | 0 | |
| | 0-,9,2+ | 3 | 7 | 4 | -2 | -1 | 2 | |
| | 0-,2+,9+ | 2 | 2 | | -3 | -2 | | |
| 3v31 | 0,3,9 | 1 | 40 | 1 | 0 | 0 | 0 | |
| | 3-,0,9+ | 3 | 11 | 2 | -2 | -1 | 2 | |
| | 0-,3-,9+ | 2 | 10 | | -3 | -2 | | |
| 3v32 | 0,4,9 | 1 | 22 | 4 | 0 | 0 | 0 | минор |
| | 0-,4,9+ | 3 | 5 | 5 | -2 | -1 | 3 | |
| | 0-,4+,9+ | 2 | 4 | | -3 | -2 | | |
| 3v33 | 0,5,9 | 1 | 22 | 5 | 0 | 0 | 0 | мажор |
| | 0-,5,9+ | 3 | 5 | 5 | -1 | -1 | 3 | |
| | 0-,5-,9+ | 2 | 4 | | -2 | -2 | | |
| 3v34 | 0,6,9 | 1 | 40 | 2 | 0 | 0 | 0 | |
| | 0-,9,6+ | 3 | 11 | 3 | -2 | -1 | 3 | |
| | 0-,6+,9+ | 2 | 10 | | -3 | -2 | | |
| 3v35 | 0,7,9 | 1 | 30 | 3 | 0 | 0 | 0 | |
| | 7-,0,9+ | 3 | 7 | 4 | -2 | -1 | 3 | |
| | 0-,7-,9+ | 2 | 2 | | -3 | -2 | | |
| 3v36 | 0,8,9 | 1 | 42 | 1 | 0 | 0 | 0 | |
| | 8-,0,9+ | 3 | 10 | 2 | -2 | -1 | 3 | |
| | 0-,8-,9+ | 2 | 8 | | -3 | -2 | | |
| 3v37 | 0,1,10 | 1 | 48 | 1 | 0 | 0 | 0 | |
| | 0-,10,1+ | 3 | 13 | 2 | -1 | -1 | 1 | |
| | 0-,1+,10+ | 2 | 12 | | -3 | -2 | | |
| 3v38 | 0,2,10 | 1 | 38 | 2 | 0 | 0 | 0 | |
| | 0-,10,2+ | 3 | 11 | 3 | -2 | -1 | 2 | |
| | 0-,2+,10+ | 2 | 8 | | -4 | -2 | | |
| 3v39 | 0,3,10 | 1 | 26 | 3 | 0 | 0 | 0 | |
| | 0-,3,10+ | 3 | 6 | 4 | -2 | -1 | 2 | |
| | 0-,3+,10+ | 2 | 2 | | -4 | -2 | | |
| 3v40 | 0,4,10 | 1 | 38 | 3 | 0 | 0 | 0 | |
| | 4-,0,10+ | 3 | 10 | 4 | -2 | -1 | 2 | |
| | 0-,4-,10+ | 2 | 6 | | -2 | -1 | | |
| 3v41 | 0,5,10 | 1 | 22 | 3 | 0 | 0 | 0 | квартовый |
| | 0-,5,10+ | 3 | 4 | 5 | -2 | -1 | 1 | |
| | 0-,5-,10+ | 2 | 4 | | -2 | -1 | | |
| 3v42 | 0,6,10 | 1 | 38 | 1 | 0 | 0 | 0 | |
| | 0-,10,6+ | 3 | 10 | 2 | -2 | -1 | 2 | |
| | 0-,6+,10+ | 2 | 6 | | -3 | -2 | | |
| 3v43 | 0,7,10 | 1 | 26 | 3 | 0 | 0 | 0 | |

| | | | | | | | |
|---|---|---|---|---|---|---|---|
| | 0-,7,10+ | 3 | 6 | 3 | -2 | -1 | 3 |
| | 0-,7-,10+ | 2 | 2 | | -3 | -2 | |
| 3v44 | 0,8,10 | 1 | 38 | 3 | 0 | 0 | 0 |
| | 8-,0,10+ | 3 | 11 | 3 | -2 | -1 | 2 |
| | 0-,8-,10+ | 2 | 8 | | -2 | -1 | |
| 3v45 | 0,9,10 | 1 | 48 | 1 | 0 | 0 | 0 |
| | 9-,0,10+ | 3 | 13 | 2 | -2 | -1 | 1 |
| | 0-,9-,10+ | 2 | 12 | | -2 | -1 | |
| 3v46 | 0,1,11 | 1 | 56 | 0 | 0 | 0 | 0 |
| | 0-,11,1+ | 3 | 17 | 1 | -1 | -1 | 1 |
| | 0-,1+,11+ | 2 | 14 | | -3 | -3 | |
| 3v47 | 0,2,11 | 1 | 48 | 1 | 0 | 0 | 0 |
| | 0-,2,11+ | 3 | 14 | 2 | -2 | -1 | 2 |
| | 0-,2+,11+ | 2 | 12 | | -3 | -2 | |
| 3v48 | 0,3,11 | 1 | 38 | 1 | 0 | 0 | 0 |
| | 0-,3,11+ | 3 | 9 | 2 | -1 | -1 | 2 |
| | 0-,3+,11+ | 2 | 8 | | -2 | -2 | |
| 3v49 | 0,4,11 | 1 | 28 | 2 | 0 | 0 | 0 |
| | 0-,4,11+ | 3 | 4 | 3 | -2 | -1 | 2 |
| | 0-,4+,11+ | 2 | 2 | | -3 | -2 | |
| 3v50 | 0,5,11 | 1 | 42 | 1 | 0 | 0 | 0 |
| | 0-,5,11+ | 3 | 11 | 2 | -2 | -1 | 2 |
| | 0-,5-,11+ | 2 | 4 | | -3 | -2 | |
| 3v51 | 0,6,11 | 1 | 42 | 1 | 0 | 0 | 0 |
| | 0-,6,11+ | 3 | 11 | 2 | -1 | -1 | 1 |
| | 0-,6+,11+ | 2 | 4 | | -2 | -2 | |
| 3v52 | 0,7,11 | 1 | 28 | 2 | 0 | 0 | 0 |
| | 0-,7,11+ | 3 | 4 | 3 | -2 | -1 | 2 |
| | 0-,7-,11+ | 2 | 2 | | -3 | -2 | |
| 3v53 | 0,8,11 | 1 | 38 | 1 | 0 | 0 | 0 |
| | 0-,8,11+ | 3 | 9 | 2 | -2 | -1 | 2 |
| | 0-,8-,11+ | 2 | 8 | | -2 | -2 | |
| 3v54 | 0,9,11 | 1 | 48 | 1 | 0 | 0 | 0 |
| | 0-,9,11+ | 3 | 14 | 2 | -1 | -1 | 2 |
| | 0-,9-,11+ | 2 | 12 | | -2 | -2 | |
| 3v55 | 0,10,11 | 1 | 56 | 0 | 0 | 0 | 0 |
| | 10-,0,11+ | 3 | 17 | 1 | -2 | -1 | 1 |
| | 0-,10-,11+ | 2 | 14 | | -3 | -2 | |